# Undetectable quantum transfer through a continuum

Jing Ping,[1] Yin Ye,[1] Xin-Qi Li,[1,2] YiJing Yan,[3] and Shmuel Gurvitz[4]

[1]*State Key Laboratory for Superlattices and Microstructures, Institute of Semiconductors,
Chinese Academy of Sciences, P.O. Box 912, Beijing 100083, China*
[2]*Department of Physics, Beijing Normal University, Beijing 100875, China*
[3]*Department of Chemistry, Hong Kong University of Science and Technology, Kowloon, Hong Kong*
[4]*Department of Particle Physics and Astrophysics,
Weizmann Institute of Science, Rehovot 76100, Israel*
(Dated: August 28, 2018)

We demonstrate that a quantum particle, initially prepared in a quantum well, can propagate through a reservoir with a continuous spectrum and reappear in a distant well without being registered in the reservoir. It is shown that such a passage through the reservoir takes place even if the latter is *continuously* monitored. We discuss a possible experimental realization of such a teleportation phenomenon in mesoscopic systems.

PACS numbers: 03.65.Yz, 42.50.-p, 73.23.-b

## I. INTRODUCTION

It is well-known that quantum motion of a wave-packet is a result of quantum interference between the wave-packet's components of different energies. Despite its interference feature, the quantum motion of a wave packet in a continuum is similar to a free motion of a classical particle. However, in some cases, the quantum motion through a continuum can be drastically different from its classical counterpart due to quantum interference on large scales.

Consider for instance a quantum well coupled to an infinite reservoir. Then a particle initially localized inside the well would eventually disappear in the reservoir. This is not surprising, since number of states in the reservoir is infinitely larger than that in the quantum well. However, if an another well is coupled to the same reservoir, as shown in Fig. 1, the situation can be quite different. It was demonstrated in Refs. [1,2] that the particle can be found in the second well with a *finite* probability (1/4 for a symmetric case) at $t \to \infty$, provided that the energy levels of both wells are aligned, i.e. $E_1 = E_2$ in Fig. 1.

This phenomenon may resemble the population trapping in the context of Quantum Optics[3]. The latter, however, is generated by two-photon resonant transitions between atomic levels through an isolated intermediate state. As a result, it would take much longer time than in the case of direct transition across the continuum, Fig. 1[2,3]. The population trapping can be increased by using the STIRAP (Stimulated Raman Adiabatic Passage) method[4] or its extension for spacial transport via a middle potential—the CTAP (Coherent Transport by Adiabatic Passage)[5].

A common signature of the CTAP-like methods is the vanishing occupation of the intermediate state, generated by the "dark state". The latter resembles the bound state embedded in the continuum[2] that generates direct transitions across the middle reservoir in Fig. 1. Thus, one can assume that the particle would not appear in the middle reservoir during the inter-well transitions, as well. However, such a bypassing of the middle reservoir would be impossible since the two wells in Fig. 1 are not directly coupled, but only through the reservoir. This implies that the particle must appear in some or another way inside the reservoir during the inter-well transitions. In order to confirm this scenario one must *continuously* monitor an appearance the particle in the reservoir. If the particle motion takes place through the reservoir, such a continuous measurement would destroy the "dark" transport.

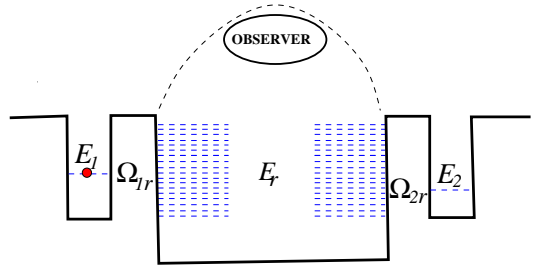

FIG. 1: (color online) A particle in two quantum wells separated by a reservoir, which is continuously monitored. $E_{1,2}$ and $E_r$ are the energy levels of the wells and of the reservoir, and $\Omega_{1(2)r}$ denote the couplings between the wells and the reservoir.

This was confirmed by investigation of CTAP[6,7], where the dark transport proceeds through an intermediate isolated state. The recent study showed that the monitoring of this state with a ballistic quantum point-contact would indeed destroy the CTAP transport[7]. However, if the two wells are separated by the reservoir, the inter-well transition is very different in its nature. We demonstrate in this paper that contrary to the expectations, the continuous monitoring of the reservoirs (Fig. 1) does not affect the transitions between the two wells. That means that the particle appears in the right well with a finite probability, without being registered in the reservoir. Moreover, we demonstrate that such an undetectable ("teleportation") phenomenon[10] characterizes all transitions through the continuum between two distant isolated states. For instance, if the particle is detected in the reservoir, it would not appear in the right well anymore.

The paper is organized as follows. In Sec. II we describe a single particle motion between two distant quantum wells, separated by the reservoir without any measurement of the particle in the reservoir[2]. A role of a bound state embedded in the continuum is elaborated. We also cast the master equations describing the unitary evolution of the system in a Lindblad form. In Sect. III we investigate continuous monitoring of the particle motion in the reservoir by applying continuous projections on the states in the wells and using the quantum trajectory approach. Numerical results illustrate main features of the undetectable quantum transfer between distant wells. In Sec. IV we suggest a possible realization of such a teleportation phenomenon in mesoscopic systems. Finally, in Sec. V we briefly summarize the work.

## II. QUANTUM PARTICLE IN TWO DISTANT WELLS SEPARATED BY A CONTINUUM

### A. Tunneling Hamiltonian

Consider two quantum wells coupled to a common reservoir with a single particle places inside one of the wells, Fig. 1. This particle can be either a boson or a fermion. For a simplicity, we assume that each of the wells contains only one level ($E_1$ and $E_2$). The reservoir states, $E_r$, are very dense (continuum) with a density of states $\varrho$. The system is described by the following tunneling Hamiltonian:

$$H = E_1 |1\rangle\langle 1| + E_2 |2\rangle\langle 2| + \sum_r E_r |r\rangle\langle r|$$
$$+ \sum_r \left( \Omega_{1r} |1\rangle\langle r| + \Omega_{2r} |2\rangle\langle r| + H.c. \right), \quad (1)$$

where $|1\rangle$, $|2\rangle$ are *localized* states of the quantum wells and $|r\rangle$ denote *extended* states of the reservoir. In the absence of a magnetic field the couplings between the reservoir and the wells, $\Omega_{1r}$ and $\Omega_{2r}$, are real.

We are interesting in the case when the couplings $\Omega$ are independent (weakly dependent) of the reservoir state $|r\rangle$, i.e. $\Omega_{1(2)r} = \Omega_{1(2)}$. These couplings are given by the overlap of the wave function of the localized state $|1, 2\rangle$ with the extended state $|r\rangle$ inside the tunneling barrier. Since the state $|r\rangle$ belongs to the continuum, it would oscillate inside the reservoir. Then the ratio $\Omega_{1r}/\Omega_{2r}$ will oscillate with a frequency $\sim E_r^{1/2} L$, where $L$ is a length of the reservoir. This creates a problem of how to make the both couplings ($\Omega_{1(2)r}$) energy independent. The problem can be avoided by coupling two quantum dots to a quantum wire at the same end, similar to the setup in Ref. [13], where two oscillators were coupled to a common reservoir at close points. Then one can make the $\Omega_{1(2)r}$ independent of the energy $E_r$ in the case of wide-band limit. Moreover, one can couple each of the dots to the reservoir through a long tunneling barrier with the energy levels of the dots close to the barrier top. Then, the coupling will not be exponentially small and the dots can be put relatively far away, one from the other[?] .

### B. Bound state embedded in the continuum

The wave function of a quantum particle in this system can be written in the most general way as

$$|\Psi(t)\rangle = b_1(t)|1\rangle + b_2(t)|2\rangle + \sum_r b_r(t)|r\rangle, \quad (2)$$

where $b_{1,2}(t)$ and $b_r(t)$ are the probability amplitudes of finding the electron in the wells or in the reservoir, respectively. These amplitudes are obtained from the time-dependent Schrödinger equation

$$i\, \partial_t |\Psi(t)\rangle = H|\Psi(t)\rangle, \quad (3)$$

where the Hamiltonian $H$ is given by Eq. (1). One easily finds that in the case of aligned levels, $E_1 = E_2$, the localized state

$$|1'\rangle = \frac{1}{\sqrt{\Omega_1^2 + \Omega_2^2}} \left( \Omega_2 |1\rangle - \Omega_1 |2\rangle \right), \quad (4)$$

is an eigenstate of the Hamiltonian (1). Indeed

$$H|1'\rangle = \frac{1}{\sqrt{\Omega_1^2 + \Omega_2^2}} \left( E_1 \Omega_2 |1\rangle - E_2 \Omega_1 |2\rangle \right) = E_1 |1'\rangle \quad (5)$$

The state $|1'\rangle$ reveals a striking example of a bound (localized) state embedded in the continuum[15]. This state is not hybridized with the reservoir states and therefore is similar to a "dark" state in quantum optics. In contrast, the localized state $|2'\rangle$, orthogonal to $|1'\rangle$

$$|2'\rangle = \frac{1}{\sqrt{\Omega_1^2 + \Omega_2^2}} \left( \Omega_1 |1\rangle + \Omega_2 |2\rangle \right), \quad (6)$$

is not an eigenstate of the Hamiltonian (1). It decays to the reservoir as $e^{-(\Gamma_1 + \Gamma_2)t}$, where $\Gamma_{1,2} = 2\pi\varrho\Omega_{1,2}^2$ is the width of the levels $E_{1,2}$ (see Ref. 2).

Since any initial state $|\Psi(0)\rangle$ inside the wells can be represented as a linear superposition of the states $|1'\rangle$ and $|2'\rangle$,

$$|\Psi(0)\rangle = \alpha_1 |1'\rangle + \alpha_2 |2'\rangle, \quad (7)$$

it would not totally decays to the continuous states of the reservoir in the limit $t \to \infty$, but survives with the probability $|\alpha_1|^2 = |\langle 1'|\Psi(0)\rangle|^2$. For instance if the particle is initially localized in the left well of Fig. 1, $|\Psi(0)\rangle = |1\rangle$, it can be found at $t \to \infty$ in the right well with a *finite* probability $|\langle 1'|1\rangle|^2 |\langle 1'|2\rangle|^2$. In the symmetric case the probability of such a transition from the left to the right well across the continuum is 1/4. The corresponding transition-time ($\sim 1/\Gamma$) would be very fast. Note that there are no localized eigenstates in the continuous spectrum if the resonant condition is not satisfied, $|E_1 - E_2| \neq 0$. In this case the state $|1'\rangle$ eventually decays to the continuum, as well.

### C. Lindblad master equations

In order to determine the time-evolution of a prepared state localized in the wells, one needs to solve the time-dependent Scrödinger equation (3) for the wave function $|\Psi(t)\rangle$, Eq. (2). It would be useful, however to use the density-matrix $\sigma_{jj'}(t)$ with $j, j' = \{1, 2\}$, instead of the wave function, defined as

$$\sigma_{11}(t) = |b_1(t)|^2, \quad \sigma_{22}(t) = |b_2(t)|^2, \quad \sigma_{12}(t) = b_1(t) b_2^*(t). \tag{8}$$

It was shown in Ref. [2] that Eq. (3) can be reduced to the master equations for this density matrix by tracing the reservoir states in the equation of motion. One finds

$$\dot\sigma_{11}(t) = -\Gamma_1 \sigma_{11}(t) - \pi \varrho\, \Omega_1 \Omega_2 \left[\sigma_{12}(t) + \sigma_{21}(t)\right] \tag{9a}$$
$$\dot\sigma_{22}(t) = -\Gamma_2 \sigma_{22}(t) - \pi \varrho\, \Omega_1 \Omega_2 \left[\sigma_{12}(t) + \sigma_{21}(t)\right] \tag{9b}$$
$$\dot\sigma_{12}(t) = i(E_2 - E_1)\sigma_{12}(t) - \pi \varrho\, \Omega_1 \Omega_2 \left[\sigma_{11}(t) + \sigma_{22}(t)\right]$$
$$- \frac{\Gamma_1 + \Gamma_2}{2} \sigma_{12}(t). \tag{9c}$$

Equations (9) can be rewritten as the Lindblad equation

$$\dot\sigma = -i[H_S, \sigma] + \Gamma_1 \mathcal{D}[a_1 + \chi a_2]\sigma. \tag{10}$$

Here $H_S = E_1 |1\rangle\langle 1| + E_2 |2\rangle\langle 2|$, $\chi = \Omega_2/\Omega_1$ and $a_1 = |0\rangle\langle 1|$, $a_2 = |0\rangle\langle 2|$, where the state $|0\rangle$ corresponds to empty quantum wells with the particle is inside the reservoir, so the corresponding density-matrix element is $\sigma_{00}(t) = 1 - \sigma_{11}(t) - \sigma_{11}(t)$. The Lindblad super-operator is defined through

$$\mathcal{D}[a]\sigma = a\sigma a^\dagger - \frac{1}{2}\{a^\dagger a, \sigma\}, \tag{11}$$

where $a \equiv a_1 + \chi a_2$.

Equations (9) (or (10)) can be solved analytically. For instance, for the symmetric case, $\Gamma_1 = \Gamma_2 = \Gamma$ and for the initial conditions corresponding to the particle in the left well, the probability of finding the particle inside the left and the right well are[1,2]

$$\sigma_{11}(t) = \frac{\Gamma^2 \cosh^2(\omega t/2) - \varepsilon^2}{\omega^2} e^{-\Gamma t}, \tag{12a}$$

$$\sigma_{22}(t) = \frac{\Gamma^2 \sinh^2(\omega t/2)}{\omega^2} e^{-\Gamma t}, \tag{12b}$$

where $\omega = \sqrt{\Gamma^2 - \varepsilon^2}$.

If follows from these equations that for aligned levels ($\varepsilon = 0$) probability of finding the particle inside the wells, $\sigma_{11}(t \to \infty) = \sigma_{22}(t \to \infty) = 1/4$. If, however, $\varepsilon \neq 0$, one finds $\sigma_{11}(t \to \infty) = \sigma_{22}(t \to \infty) = 0$. Nevertheless for small displacement $\varepsilon \lesssim \Gamma$, the particle spends long time ($\sim \Gamma/\varepsilon^2$) inside the right well before it decays to the reservoir[2]. Thus the effect is still exists, even so the condition of aligned levels is not precisely fulfilled.

## III. PARTICLE TRANSFER THROUGH THE RESERVOIR UNDER CONTINUOUS NULL-RESULT MEASUREMENT

It is natural to assume that a particle, initially localized in the left well, arrives to the right well from the reservoir. In particular, this is expected from the Hamiltonian (1) which couples two wells only through the reservoir. In order to check this point, we have to monitor the reservoir continuously on the time-interval $(0, t)$. For this reason we insert the detector ("observer"), Fig. 1, that registers the particle in the reservoir at each time-intervals $\Delta t = t/n$, choosing only the events where no particle is detected in the reservoir (the null-result measurements). Then in the limit of *continuous* null-result measurement, $n \to \infty$, one would anticipate that the particle remains *locked* in the left well and therefore would never appear in the right well.

Indeed, consider the initial state $|\Psi(0)\rangle = |1\rangle$, corresponding to the particle localizes in the left well. We represent it as a superposition of the states $|1'\rangle$ and $|2'\rangle$, Eq. (7) with $\alpha_{1,2} = \Omega_{2,1}/\sqrt{\Omega_1^2 + \Omega_2^2}$. In the case of aligned levels, $E_1 = E_2 = 0$, the state $|1'\rangle$ is an eigenstate of the Hamiltonian $H$ (see Eq. (5)) with the energy $E = 0$. Then after the time interval $\Delta t$ the wave function becomes

$$|\Psi(\Delta t)\rangle = e^{-iH\Delta t}|\Psi(0)\rangle$$
$$= \alpha_1 |1'\rangle + \alpha_2 \left[1 - iH\Delta t - \frac{1}{2} H^2 (\Delta t)^2 + \cdots\right]|2'\rangle \tag{13}$$

The null-result measurement in the reservoir implies that the particle is inside the wells. Therefore the wave function is projected to the subspace of the two wells, $|\Psi(t)\rangle \to \mathcal{P}|\Psi(t)\rangle$, where

$$\mathcal{P} = \frac{1}{N}\big(|1'\rangle\langle 1'| + |2'\rangle\langle 2'|\big) \tag{14}$$

and $N$ is a normalization factor. Projecting Eq. (13) on the dots states and taking into account that $\langle 1'|H|2'\rangle = 0$ and $\langle 2'|H|2'\rangle = 0$, we obtain

$$|\Psi_1\rangle = \mathcal{P}|\Psi(\Delta t)\rangle = \frac{1}{N_1}\Big[\alpha_1 |1'\rangle + \alpha_2 \big(1 - C(\Delta t)^2\big)|2'\rangle\Big] \tag{15}$$

where $C = \sum_r \Omega_r^2/2$ and $N_1^2 = 1 - 2\alpha_2^2 C(\Delta t)^2$. After $n$ projections, corresponding to the $n$ subsequent null-result measurements, where $n = t/\Delta t$ we find

$$|\Psi_n\rangle = \frac{1}{N_n}\Big[\alpha_1|1'\rangle + \alpha_2\big[1 - C(\Delta t)^2\big]^n|2'\rangle\Big] \quad (16)$$

where $N_n = \sqrt{1 + 2n\,\alpha_2^2 C\,(\Delta t)^2}$. Thus in the limit of the continuous null-result measurement, $\Delta t \to 0$ and $t$=const., we obtain

$$|\Psi_n\rangle \to \alpha_1|1'\rangle + \alpha_2|2'\rangle \equiv |1\rangle. \quad (17)$$

Thus no transitions between the separated dots are expected in the case of continuous monitoring of the reservoir. Nevertheless the following analysis shows that in spite of the above arguments the particle can make an undetectable transition between the wells.

### A. Quantum trajectory method

In fact there exists a well-developed theory of continuous quantum measurement, describing the measurement-results conditioned state evolution. The resulting evolution is governed by the quantum trajectory equation[16]. In our case this equation reads:

$$d\rho = dN(t)\,\mathcal{G}\,[a]\,\rho - dt\,\mathcal{H}\,\big[iH_S + \Gamma_1 a^\dagger a/2\big]\rho. \quad (18)$$

Here we use $\rho \equiv \rho(t)$ to denote the conditional density-matrix, rather than unconditional one, $\sigma(t)$ in Eqs. (9). The super-operators are defined as[16] $\mathcal{G}[a]\rho = a\rho a^\dagger/\text{Tr}[a\rho a^\dagger] - \rho$, and $\mathcal{H}[x]\rho = x\rho + \rho x^\dagger - \langle x + x^\dagger\rangle\rho$ where $\langle\cdots\rangle \equiv \text{Tr}[(\cdots)\rho]$. $dN(t) = 0$ or $1$, being the measurement record of particle number detected in the reservoir during the time interval $(t, t+dt)$.

The above Eq. (18) is associated with an evolution based on a given measurement record. Consider the evolution under condition of the null-result measurement. For continuous measurement it corresponds to $dN(t) = 0$ in Eq. (18). Conditioned on this requirement, the state evolution is given by

$$\dot\rho = -\mathcal{H}\,\big[iH_S + \Gamma_1 a^\dagger a/2\big]\rho$$
$$= -i[H_S,\rho] + \Gamma_1\Big(\text{Tr}[a^\dagger a\rho]\,\rho - \frac{1}{2}\{a^\dagger a,\rho\}\Big). \quad (19)$$

**It should be pointed out that the Markovian dynamics of the reservoir plays a crucial role in obtaining Eqs. (18), (19). Indeed, in the case of non-Markovian dynamics, a measurement on the reservoir would affect the memory stored in it, and therefore modify a measurement unravelling evolution of the system[17].**

Equation (19) is actually a *nonlinear* equation. Nevertheless its solution can be written in a simple way by introducing an effective evolution operator $\mathcal{U}_{\text{eff}}(t,0)$, as follows

$$\rho(t) = \mathcal{U}_{\text{eff}}(t,0)\rho(0)\mathcal{U}_{\text{eff}}^\dagger(t,0)/||\cdot||, \quad (20)$$

where $\mathcal{U}_{\text{eff}}(t,0) = \exp\big[-i\big(H_S - i\frac{\Gamma_1}{2}a^\dagger a\big)t\big]$. Here $||\cdot||$ denotes the trace of the numerator that normalizes the state, reflecting a non-unitary of $\mathcal{U}_{\text{eff}}(t,0)$.

Equation (20) can be easily solved in the basis of the states $|1'\rangle$ and $|2'\rangle$, Eqs. (4), (6). The solution has a particularly simple form for the symmetric case: $\Gamma_1 = \Gamma_2 = \Gamma$. One finds from Eq. (20) that

$$\rho_{11}(t) = \frac{\cosh\omega t + 1 - 2\varepsilon^2/\Gamma^2}{2\big(\cosh\omega t - \varepsilon^2/\Gamma^2\big)} \quad (21a)$$

$$\rho_{22}(t) = \frac{\cosh\omega t - 1}{2\big(\cosh\omega t - \varepsilon^2/\Gamma^2\big)} \quad (21b)$$

(c.f. with Eqs. (12)). Thus, contrary to the expectations, the particle is not locked in the left well under the *continuous* null-result measurement condition in the reservoir. It can be found in the right well with a *finite* probability. This implies that the particle jumps directly to the right well without passing the reservoir, although the wells can be largely separated in space. Since such a jump cannot take place via the reservoir states, it proceeds through the bound state $|1'\rangle$, Eq. (4), embedded in the continuum. This state is a part of the total Hamiltonian spectrum. However it does not belong to the reservoir states and therefore the transport via this state cannot be detected by continuously monitoring the reservoir.

Comparing $\rho_{11}(t)$ and $\rho_{22}(t)$ with $\sigma_{11}(t)$ and $\sigma_{22}(t)$, given by Eqs. (12), one finds that the both quantities are quite different. For instance, $\rho_{11} = \rho_{22} = 1/2$ in the asymptotic limit $(t \to \infty)$ for any $\varepsilon$, whereas in the same limit $\sigma_{11} = \sigma_{22} = 1/4$ for $\varepsilon = 0$ and $\sigma_{11} = \sigma_{22} = 0$ for any $\varepsilon \neq 0$. This is not surprising since $\rho(t)$ is the conditional probability, subjected to the requirement that only events with no particle in the reservoir are counted. On the other hand, $\sigma(t)$ is the unconditional density matrix describing all events, including those when the particle disappears in the reservoir. As a result the probability of finding the particle inside the wells, $P_0(t) = \sigma_{11}(t) + \sigma_{22}(t)$, is smaller than one. Indeed, $P_0(t \to \infty) = 1/2$, Eq. (12). The conditional density matrix, $\sigma_{jj'}^{(c)}(t)$, for finding the particle in each of the wells at time $t$ is therefore

$$\sigma_{jj'}^{(c)}(t) = \frac{\sigma_{jj'}(t)}{\text{Tr}[\sigma(t)]} = \frac{\sigma_{jj'}(t)}{P_0(t)}, \quad \text{for } j, j' = \{1, 2\}. \quad (22)$$

In contrast with $\rho(t)$, which implies that no particle is found in the reservoir on a *whole* time-interval time $(0, t)$, the conditional probability $\sigma_{jj'}^{(c)}(t)$, given by Eqs. (9), (10), is much less restrictive. It only implies that no particle in the reservoir is found at time $t$. However at any other time $t'$ inside the interval $(0, t)$, the particle can appear in the reservoir. As a result, one would expect that the conditional probability $\sigma_{jj'}^{(c)}(t)$, obtained from the Schrödinger evolution would exceed $\rho(t)$ obtained from the quantum trajectory equation (19). However,

it follows from Eqs. (12) and (21) that

$$\rho_{jj'}(t) = \sigma^{(c)}_{jj'}(t). \quad (23)$$

Although the l.h.s. and r.h.s. of Eq. (23) are related to different systems (with and without the detector), this equality implies a striking picture: it looks like that in *all* transitions between the two distant wells (measured or not), the particle does not appear in the reservoir.

In fact, Eq. (23) can be obtained from rather general arguments. Let us consider Eqs. (3) describing the evolution of the total wave function, $|\Psi(t)\rangle$. The null-result measurement in the reservoir corresponds to $|\Psi(t)\rangle \to \mathcal{P}|\Psi(t)\rangle$, where the projection operator $\mathcal{P}$ is given by Eq. (14). It is very crucial for our arguments that **in the case of wide band limit (Markovian dynamics)** the Schrödinger equation (3) can be reduced to the system of closed linear equations (9) for the reduced density matrix $\sigma(t)$ in the subspace of the two wells. In this subspace the projector $\mathcal{P}$, Eq. (14), is proportional to the *unit* matrix, up to the overall normalization. Therefore any repeated $n$ null-result measurements correspond to $n$ subsequent projections of the density matrix, $\mathcal{P}\sigma(t)\mathcal{P}$, affecting its time development given by Eqs. (9), by the normalization factor only. As a result, the conditional density matrix, subjected to $n$ null-result measurement, $\rho(t)$ is given by

$$\rho(t) = \frac{1}{\mathcal{N}_1} \cdots \frac{1}{\mathcal{N}_n} \sigma(t) = \frac{\sigma(t)}{\text{Tr}[\sigma(t)]}. \quad (24)$$

which coincides with Eq. (23).

### B. Undetectable propagation under continuous monitoring of the reservoir

We thus found that the continuous null-result monitoring of the reservoir does not prevent the particle's jump between two distant wells. This conclusion has been based on Eq. (12) obtained from the Schrödinger equation, and the evaluation of conditional probabilities, subjected to the continuous null-result measurement, Eq. (24). The question arises how this result can be accommodated with the previous arguments, based on the Zeno-type effect, and leading to the opposite conclusion, Eq. (17).

In order to understand this contradiction we compare the Scrödinger evolution for probability of finding the particle inside the dots, $p_0(t)$, for small time $t = \Delta t$, given by Eqs. (12) with that given by Eq. (13). Consider the aligned levels ($\varepsilon = 0$). Using Eqs. (12) we find

$$P_0(\Delta t) = 1 - \Gamma \Delta t + \Gamma(\Delta t)^2 + \cdots \quad (25)$$

Thus $1 - P_0(\Delta t) \propto \Delta t$. On the other hand, it follows from Eq. (13) that $1 - P_0(\Delta t) \propto (\Delta t)^2$. Indeed,

$$P_0(\Delta t) = 1 - |\alpha_2|^2 \langle 2'|H^2|2'\rangle(\Delta t)^2 + \cdots \quad (26)$$

Although the latter result looks rather general, it is based on an assumption that the evolution operator can be expand in powers of $t$, Eq. (13). This assumption does not always hold[18], as for instance in the case of a flat density of the reservoir states in the Hamiltonian (1). In contrast, Eq. (12), has been obtained from the Schrödinger equation[2] without expanding the corresponding evolution operator in powers of $t$. Therefore it holds for any small time-interval.

In fact, the flat density of state **(the wide band limit)** is very important requirement for obtaining the Lindblad master equation (10) from the Schrödinger equation. This requirement characterizes the Markovian reservoirs. If the reservoir is not Markovian, Eq. (10) would be modified. **Then the continuous monitoring of the reservoir could affect the memory stored in it[17]. This in turn, could affect the subsequent evolution of the system interacting with the reservoir.** The case of non-Markovian reservoirs needs a special investigation.

### C. Numerical analysis

Consider first the conditional density matrix $\rho(t)$ for the case of the same level-widths $\Gamma_1 = \Gamma_2 = \Gamma$. The corresponding couplings $\Omega_{1,2}$, however, can be of opposite sign, $\Omega_1 = \pm\Omega_2$, depending on the relative parity of the well states, $\eta = \Omega_1/\Omega_2 = \pm 1$. Conditional probabilities for occupation of the left and right wells, $\rho_{11}(t)$ and $\rho_{22}(t)$, Eqs. (21), are displayed in Fig. 2 for the initial condition corresponding to the occupied left well. Both are independent of $\eta$. The latter affects only the off-diagonal term, as illustrated in the inset of Fig. 2. The solid curves correspond to aligned levels, $\varepsilon = 0$ and the dashed curves to displaced levels, $\epsilon \neq 0$. In both cases the occupation probabilities reach the same asymptotic limit, and the effect of level displacement is unessential. This is drastically different from the asymptotic behavior of the unconditional density matrix, $\sigma(t)$, Eq. (12), that vanishes in the asymptotic limit for any finite $\varepsilon$, no matter how small it is.

The reason for such a different behavior is quite clear. Indeed, probability of an event where the particle is *not found* inside the reservoir becomes very small for $\varepsilon \neq 0$ and large $t$, since the unconditional density matrix, $\sigma(t)$ vanishes at $t \to \infty$. However, if such an event takes place at time $t$, it implies that the particle stays inside the dots in the time interval $(t, t + \Delta t)$ with probability one for $\Delta t \to 0$. Therefore the *conditional probability* of finding the particle in each dot is not vanishes at $t \to \infty$ for $\epsilon \neq 0$. It reaches the value of $1/2$ for symmetric dots, as shown in Fig. 2.

The conditional occupation probabilities in the steady state, $\bar{\rho} = \rho(t \to \infty)$, as a function of the coupling asymmetry $\chi = \Omega_2/\Omega_1$, are shown in Fig. 3 for aligned and misaligned levels. As in Fig. 2 the level misalignment ($\varepsilon$) have minor effects on the stationary occupa-tion probabilities, in contrast with the unconditional probabilities[2]. However, similar to unconditional probabilities, the asymmetry of coupling between the continuum and the well strongly affects both the transient process and the stationary state in an unexpected way. Indeed the unconditional probability of transfer from the left to the right well, Eqs. (12), increases, when the coupling of the right well with the reservoir, $\Omega_2$, decreases! This peculiar behavior has been explained in Ref. 2 in terms of quantum interference at large scale. In the case of conditional probabilities the same behavior of transition probabilities, displayed in Fig. 3, can be understood by an information-theoretic interpretation. Consider, for instance, $\Gamma_2 < \Gamma_1$, corresponding to $\Omega_2/\Omega_1 < 1$. The null-result of measurement in the reservoir indicates that probability of finding the electron in the right dot is larger than that in the left dot. In the opposite case, $\Gamma_1 < \Gamma_2$, the electron is more likely to be found in the left dot, conditioned on the same null result of measurement.

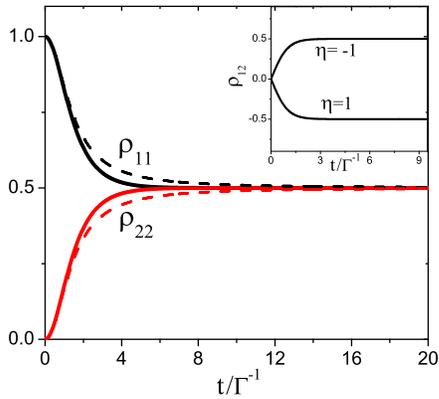

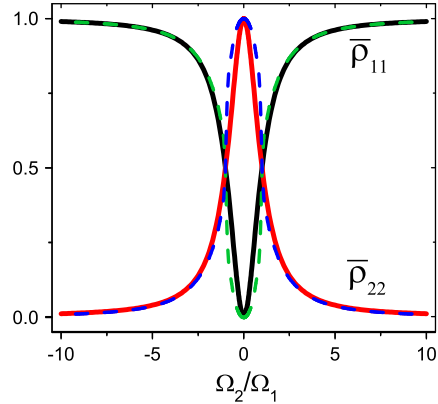

FIG. 2: (color online) Occupation of the wells, conditioned on the null result of continuous measurement in the reservoir for $\Gamma_1 = \Gamma_2 = \Gamma$. The solid lines correspond to $E_1 = E_2$, while the dashed lines to $E_1 - E_2 = \pm 0.5\Gamma$. Inset: the off-diagonal element of the density matrix for $\eta = \pm 1$ and $E_1 = E_2$.

FIG. 3: (color online) Steady-state occupation probabilities as a function of the coupling asymmetry. Similar to Fig. 2, the solid curves correspond to $E_1 = E_2$, while the dashed curves correspond to $E_1 - E_2 = \pm 0.5\Gamma$.

We also like to mention that the conditional probability $\bar\rho$ at $\chi = 0$ implies the following order of limits: $\bar\rho = \lim_{\chi \to 0} \lim_{t \to \infty} \bar\rho$. The inverse order would result in $\bar\rho_{11} = 1$ and $\bar\rho_{22} = 1$ at $\chi = 0$. Consequently, $\bar\rho$ displays a nonanalytic behavior at $\chi = \Omega_2/\Omega_1 = 0$, as shown in Fig. 3. Note also that one cannot change a sign of $\chi$ by a modulation of the barrier height. One needs to change the quantum well parameters in such a way that the corresponding wave function obtains an additional node[19].

## IV. ELECTRON TELEPORTATION TRANSFER IN MESOSCOPIC SYSTEMS

An experimental realization of continuous monitoring of the reservoir, Fig. 1, would be a rather complicated problem, since the detector ("observer") can distort the reservoir states. This in turn, could effect the electron transfer across the reservoir. One can apply, however, an alternative procedure by monitoring continuously the quantum wells, instead the reservoir.

An example of such continuous measurement in mesoscopic system is shown in Fig. 4. A non-invasive monitoring of the dots is provided by a point contact (PC) in the near proximity to two dots. If the electron occupies one of the dots, its electric field diminishes the "opening" of the PC, thus affecting the PC current $I$. The latter increases whenever the electron tunnels to the reservoir, The current decreases again if the electron returns to the dots. By placing the PC symmetrically with respect to the dots, as shown in Fig. 4, one cannot distinguish which of the dots is occupies. As a result, such a measurement would not affect the electron dynamics inside the dots, providing therefore a non-invasive monitoring of the electron transitions to the reservoir.

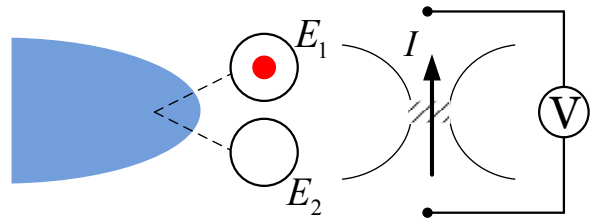

FIG. 4: (color online) Two dots, coupled with a common reservoir, are continuously monitored by a point-contact detector. The current $I$ drops down when the electron tunnels to the reservoir.

The two dots in Fig. 4 are coupled to a common reservoir in a manner explained in Sec. IIA. In this way the system can be described by the tunneling Hamiltonian (1) with weakly energy-dependent tunneling cou-plings $\Omega_{1(2)r}$. In this case, as we demonstrated above, the particle initially localized in one of the dots could appear in the second dot, without been detected in the reservoir, despite its continuous monitoring. Thus the teleportation-type phenomena could be observed in the setup of Fig. 1. The latter can be realized by using the available state-of-the-art technology.

## V. SUMMARY

In this paper we investigated the dynamics of a particle transfer between discrete states in two distant wells, separated by a reservoir with continuous spectrum. For this purpose we introduced continuous monitoring of the reservoir by an external observer (device), which registers a particle in the reservoir. An analysis of quantum motion under continuous observation can be performed by applying the quantum trajectory approach. In particular, we considered only the events where the particle never appears in the reservoir (null-result continuous measurement). Quite unexpectedly, despite such a strict restriction, we found that the particle still appears in the second distant well, making its motion in the reservoir totally undetectable (we refer to this phenomenon as "teleportation"[10]).

Such a transition through the reservoir takes place through a *localized eigenstate* (bound state) embedded in the continuum. Although this state belongs to the continuum spectrum, its wave function is not extended into the reservoir and therefore is undetectable by monitoring the reservoir. Nevertheless it can mediate quantum transitions between distant wells.

In fact, the localized eigenstate embedded in continuum is analogous to a "dark" state in atomic system. The difference is that our state represents linear superposition of two components, largely separated in space. It becomes an eigenstate embedded in the continuum spectrum due to cancelation between different components of the eigenfunction in the continuum, thus revealing distracting quantum interference effect on large scales.

One expects, however, that the quantum interference would be destroyed if the motion of quantum particle is continuously monitored by an outside observer. A well-known example is an observation of quantum motion of a single particle in two slit experiment. This argument could imply that the quantum transfer through continuum between two isolated states would be destroyed either, whenever the particle's motion in the reservoir is continuously observed. In particular, one anticipates that the continuous null-result measurement that never register the particle in the reservoir, prevents its transfer trough continuum between distant wells. The above argument makes our results presented in this paper even more surprising, since the particle still penetrates through the reservoir, even it is *not* registered in it.

**Acknowledgments**

We acknowledge to G. Huang for stimulating discussions. S.G. acknowledges the Department of Physics, Beijing Normal University, and the State Key Laboratory for Superlattices and Microstructures, Institute of Semiconductors, Chinese Academy of Sciences for supporting his visit. Meanwhile, X.Q.Li acknowledges the Department of Chemistry, Hong Kong University of Science and Technology for supporting his visit. This work was supported by the Israel Science Foundation under grant No. 711091, and NNSF of China under grants No. 101202101 & 10874176, and the Major State Basic Research Project of China under grants No. 2011CB808502 & 2012CB932704.